# Some aspects of epitaxial thin film growth

A.S.Bhattacharyya[1, 2*], Rishdeo Kumar [1], Vikrant Raj [1] , S.Shrinidhi[1],
Shrishti Suman[1], Asmita.Shah[1], Ramgiri P. Kumar[1]

[1]Centre for Nanotechnology and [2] Centre of Excellence in Green and Efficient
Energy Technology (CoE-GEET)
Central University of Jharkhand, Ranchi – 835 205, India,

*e-mail: 2006asb@gmail.com ; arnab.bhattacharya@cuj.ac.in

The article consists of four sections all dealing with the computational modeling of the sputtering process. The first section deals with the difference in Bismuth atomic layer deposition at different polar angle and ion flounce. In the second section,  atomic layer deposition was studied as a function of incident Argon ion energy at different pressure conditions. A curve analogous to Bragg curve was obtained .In the third section TiOx films was developed by simulations. The sputtering parameters were varied to get different atomic layers. The variation of coverage with sticking coefficient was shown.  The last section deals with sputter based deposition of TiN films. The rate of change of partial sputtering yield with coverage was considered. The deposition pressure and time were varied to get films of different thickness. All the sections are stored separately in other repositories

Keywords : Sputtering, modeling, epitaxial, thin film

# 1. Bismuth thin films: polar angle and ion fluence

Bi has shown major physical phenomena in the past like de Haas-van Alphen effect [1], quantum size, confinement effect [2], quantum linear magnetoresistance [3], peculiar superconductivity [4], and possibly fractional, quantum Hall effect [5]

We started with our established model of sputtering and deposited atomic layers of Bi as a function of deposition time and current density. The sputtering rate i.e the amount of target materials sputtered per unit time is given by $zt = M/(rNAe) S jp$ .where M is molar weight of the target [kg/mol]; r is the density of the material [kg/m³]; NA is the 6.02 x 10²⁶ 1/kmol; (Avogadro number); e is the 1.6x 10-19 As (electron charge); S is the sputtering yield (atom/ion) and jp: primary ion current density [A/m²]. This rate divided by atomic diameter gives us the rate of sputtering in terms of atomic layers (AL) per second [7, 8]

The input parameters were molar weight of Bi (209 g/mol), its atomic radius (1.67 Å) and density (9.8 g /cm³) [7, 8]. The results showed a linear variation of atomic layers with deposition time and a higher rate of deposition with increase in current density as shown in Fig 1(a) with the highest and lowest slope plots for j = 25 and 1 mA / cm² .

The computational model gave us 250 (25 ×10) values of atomic layers for 10 values of deposition time ( t = 1to 10s) and 25 values of current density ( j = 1 to 25 mA / cm²) the distribution of which is shown in Fig 1(b). The plot can be seen to be a combination of 25 similar shaped plots each corresponding to a particular current density value. The plots were found to be steeper for higher current density values suggesting higher rates of deposition as also found previously. The number of atomic layers multiplied by the atomic diameter gives us the film thickness. The thickness varied from few nm to 500 nm (Fig 2).

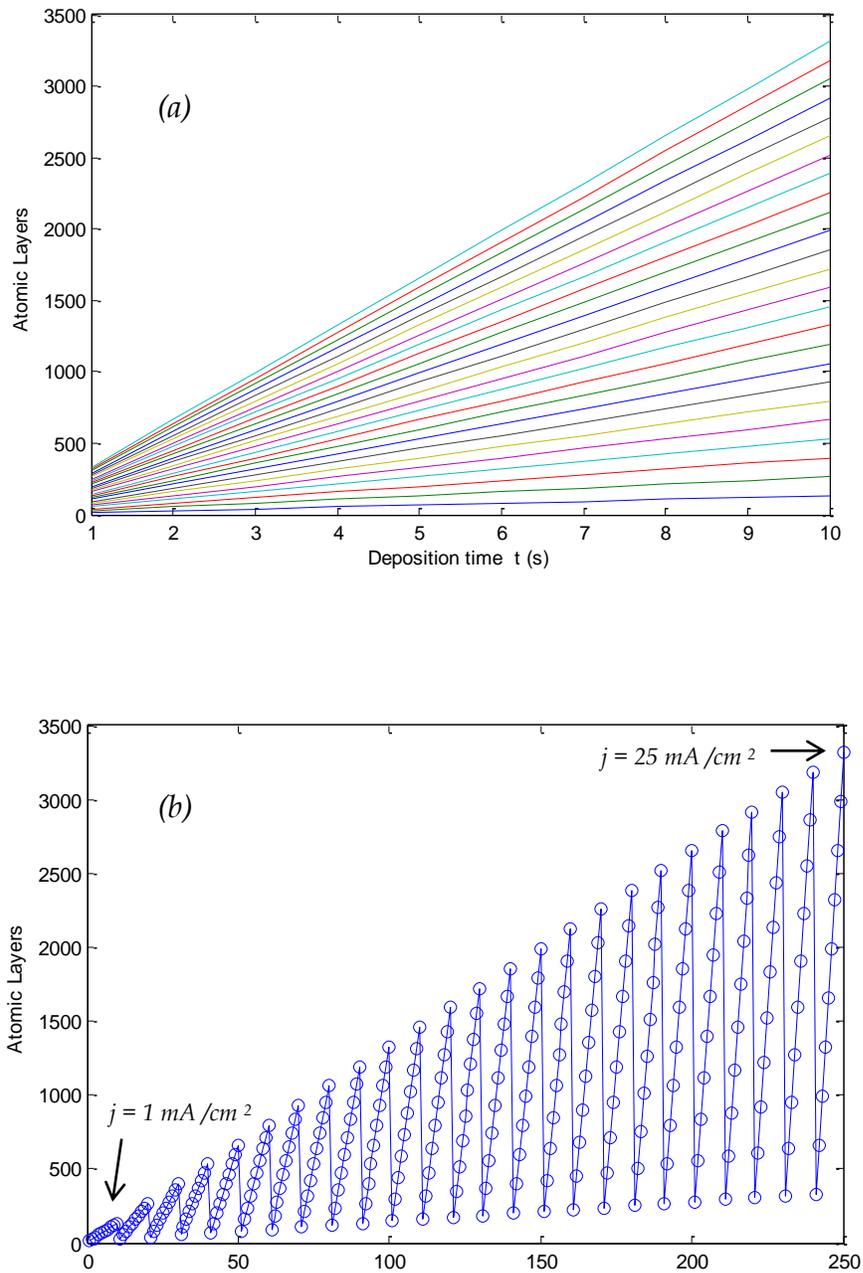

Fig 1: Bi atomic layers deposited as a function of deposition time and current density

Deoli et. Al have shown the ion fluence dependence of the total sputtering yield and differential angular sputtering yield of Bi. An attempt was made to computationally mock the sputtering process and deposit Bismuth films and study the effect of polar angle and ion fluence. Experimentally it has been shown that the differential and total sputtering yield for Bi by normally incident 50 keV Ar+ ions increases with increasing ion fluence for the measured fluence range [6]. We tried to match the experimental results and moved a step further showing the variation of Atomic Layer deposition with polar angle.

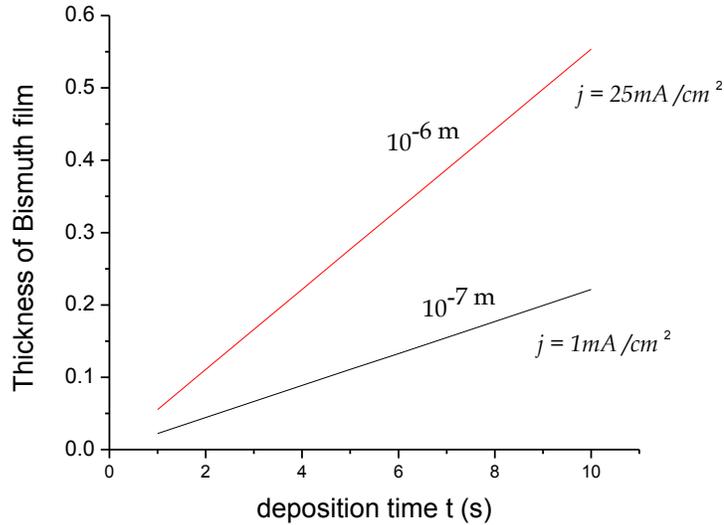

Fig 2: Deposition rate at two different current densities

In our model we then used the differential sputtering yield instead of the sputtering yield and observed specific polar angles giving higher deposition rate (Fig 3). The angles were 0 degree and 60 degree. Sputtering peaks around 60 degree which is an established fact, however the angle 0 degree can be considered as error as there is possibility of the freed atoms to be knocked deeper in to the material

Bi thin films deposited by MBE showed variation in electrical conductance with thickness and temperature. A unique feature of showing insulating properties in the bulk whereas conducting properties called Topological insulators (TI) are helpful for spintronic devices. [9]. Bi thin films are interesting examples where a topologically trivial system becomes non-trivial solely due to the reduction of thickness [10]. The topological states have been also investigated using first principle calculations [11].

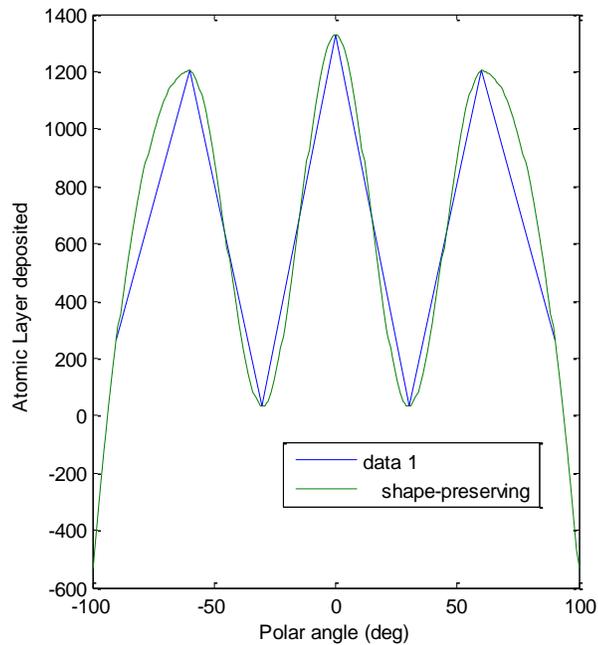

Fig 3 : Atomic layers of Bi at different polar angle

Bismuth Selenide ($Bi_2Se_3$) and Bismuth Telluride ($Bi_2Te_3$) are very promising Tis which have sown oscillatory magnetoresistance due to Shubnikov-de Haas Effect [12]. Combining TIs and superconductors give Marjorana Fermions which possess both particle and antiparticle properties. They have application in Quantum Computing [13]. This section is stored in a repository [14].

## 2. Bragg Curve in Sputtering

In our previous publication we have shown a model developed for the sputter based epitaxial growth of Cu [7, 8]. In this communication we are extending the model to study the atomic layer deposition with incident argon energy at different deposition pressure. The sputtering yields of various materials for incident Ar ions are given in Table 1[15]. The sputtering yield is found to increase with ion energy as shown in fig 1 which is quite obvious as more is the energy more will be the impact leading to more ejection of atoms from the target. An inert gas usually argon is introduced in an evacuated chamber and ionized. Plasma is formed in the process of ionization. The atomic density in plasma is usually of the order of $10^{-18}$ per $m^3$.

Table 1: Sputtering yields of different materials with Ar ion energy [15, 16]

| E keV | Al | Si | Ge | Ti | Fe | Ag | W | Au |
|---|---|---|---|---|---|---|---|---|
| 0.2 | 0.3 | 0.2 | 0.5 | 0.1 | 0.2 | 1.0 | 0.1 | 0.7 |
| 0.5 | 0.9 | 0.6 | 1.0 | 0.5 | 0.5 | 3.0 | 0.6 | 2.0 |
| 1.0 | 1.8 | 1.0 | 1.4 | 1.0 | 1.0 | 5.0 | 1.0 | 3.5 |
| 2.0 | 2.5 | 1.2 | 2.0 | 1.3 | 2.5 | 6.5 | 1.5 | 5.5 |
| 5.0 | 3.1 | 1.4 | 2.6 | 1.5 | 3.0 | 8.0 | 2.0 | 7.5 |
| 10 | 3.7 | 1.5 | 3.2 | 1.8 | 3.5 | 10.0 | 2.5 | 8.5 |

The sputtering yield S is again a function of the incident Ar ion energy E. We start with the metal tungsten (W) which has molar weight of 184g/mol, density of 12.96 g/cm³ and radius 0.137 nm. The S vs E plots for W were fitted with a quadratic polynomial as to get a relation between S and E. From fitting we obtain a relationship between S and E as follows S = - 0.033E² + 0.55E + 0.3   where we neglect the intercept value 0.3. So we replace S with E in the sputtering rate equation [16].  In the model, the argon ion energy was varied from 1 to 10 keV. The deposition pressure P was varied from 10⁻¹ Torr to 10⁻⁶ Torr in terms of 10⁻ⁿ where n is the pressure index which varied from 1 to 6. The details of the modeling with explanation on the parameters are given in our previous publication [7, 8].

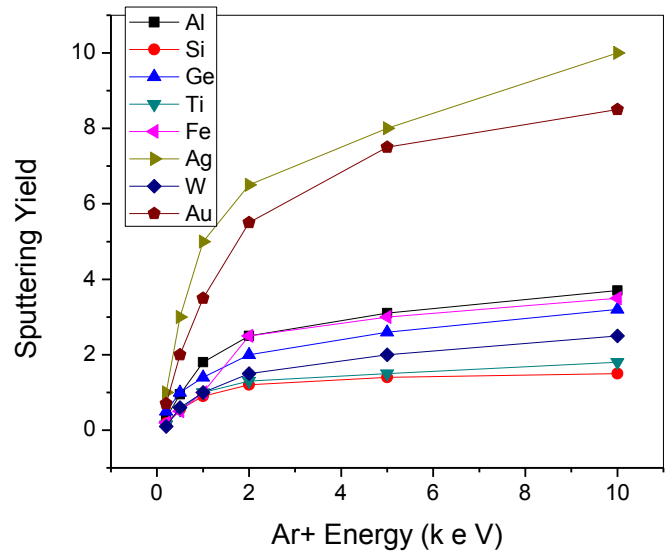

Fig 4: Sputtering yield vs. Ar ion energy [15]

The atomic layers of tungsten sputtered (AL W) and the atomic layers deposited on the substrate (ALd W) with increase in Ar ion energy are plotted in fig 5. There is not much of a difference between AL W and ALd W at lower pressures (higher vacuum) of $10^{-6}$ and $10^{-5}$ Torr. On decreasing the pressure to $10^{-3}$ and $10^{-4}$ Torr, more prominent difference is observed. As we decrease the pressure the difference increases. The reason behind this observation is the mean free path of the adatoms at lower pressure which results in their less collisions with the gas atoms which has been explained previously [7, 8].The film thickness naturally follows the same trend as shown in fig 6.

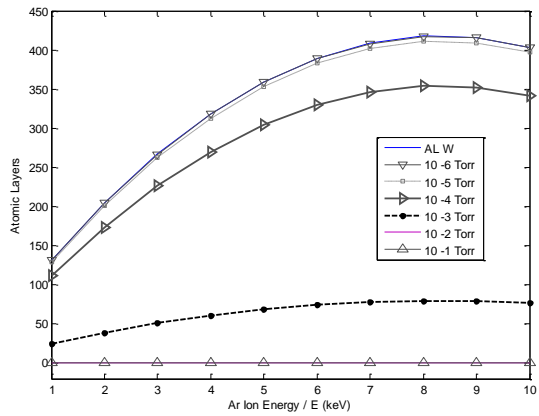

Fig 5: Atomic layers of tungsten sputtered (AL W) and atomic layers deposited at different ion energies and deposition pressure [8]

The difference in atomic layers sputtered and deposited denoted by D with respect to the pressure index at different energies is given in fig 4. The difference was higher at higher energies but the rate of change of difference showed a decrease with increase in energy.

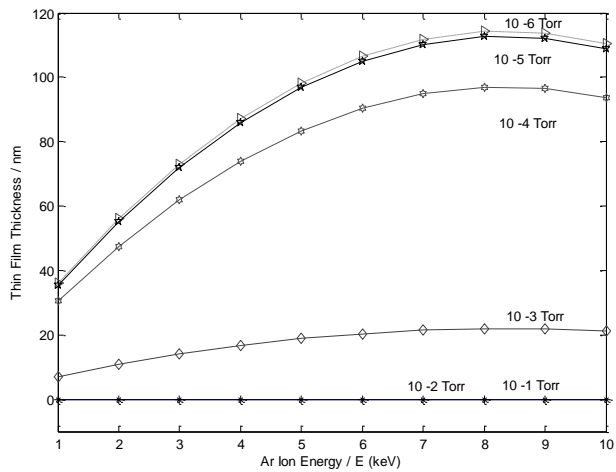

Fig 6: The film thickness of tungsten deposited at different ion energies and deposition pressure [8]

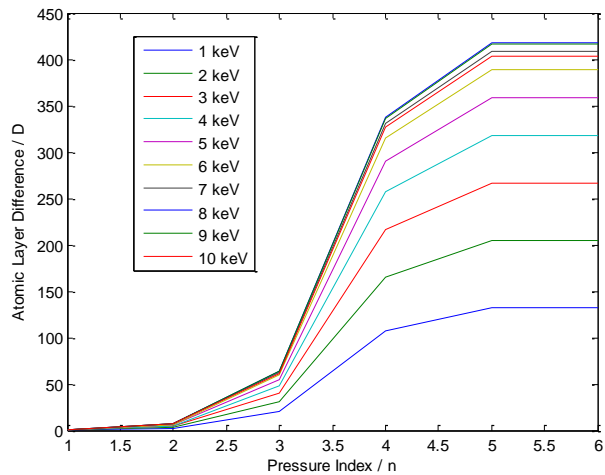

Fig 7: The difference in atomic layers sputtered and deposited w.r.t deposition pressure (index n) at different Argon ion energies

The rate of change of energy with thickness at different ion energies is given in fig 8. The rate of change in energy with respect to thickness difference on the other hand given in fig 9 resembles the Bragg curve for ion radiation. By drawing an analogy with the Bragg curve, we can treat $(dE/dD_{Th})$ having using keV/cm as the stopping potential S $(D_{Th})$. So from fig 8 we can say that the Ar ions is effective for approximately 17 nm in case of 10-3Torr and 90nm and 110nm for $10^{-5}$ and $10^{-6}$ Torr pressure respectively. Hence an increase in incident ion effectiveness increases with decrease in pressure. This section is stored in a repsitory[17]

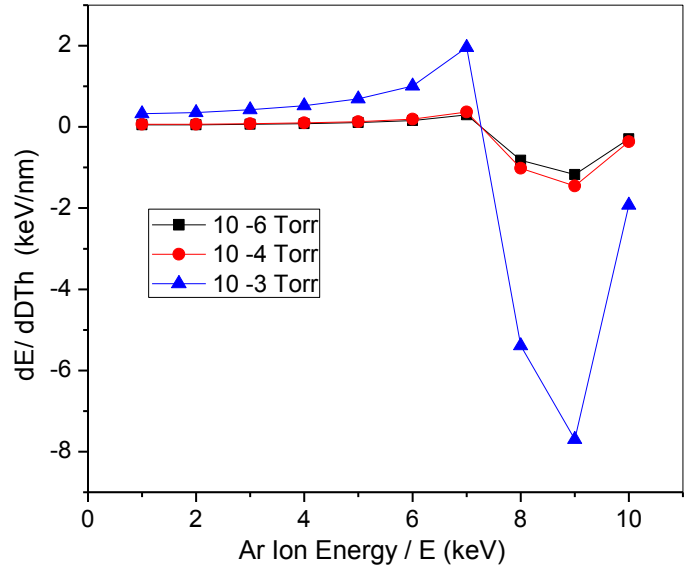

Fig 8: The rate of change of Thickness with Ar Ion energy at different deposition pressures

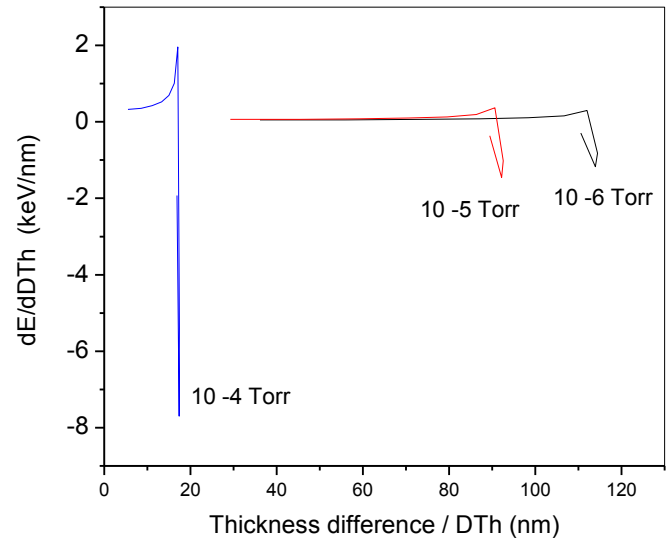

Fig 9 : Curve analogous to Bragg curve showing the rate of change of Ar ion energy with thin film thickness difference at different deposition pressure with stopping potential $S(D_{Th})=dE/dD_{Th}$

## 3. Compound TiO$_2$ film growth dynamics

Sputtering is an effective means of depositing thin films. Compound films of most of the materials can be developed by reactive sputtering. The sputtering rate of the compound material formed by reaction of gas introduced in the sputtering chamber with the target is much less than that of target itself. The target is also called poisoned target. The hysteresis and effect of reactive gas in target poisoning is published in ref [18]. It has been shown that sputtering should be done in the transition region between elemental and metallic states for best properties of compound film. The hysteresis effect can be somewhat reduced by increasing the pumping speed of the system [19]. However increasing the pumping speed is not cost effective and control of the reactive gas flow is more convenient [20-22].There is also a hysteresis free operation area if the sputtering area is reduced to less than a critical size such that the compound forms on the substrate before the target [23].

A model has been proposed to study the reactive sputtering process [7]. A model has also been developed for the epitaxial growth of metals which is also based on the atoms getting sputtered out from the target surface and ultimately being deposited on the substrate to form the film [7, 15]. Considering a compound film thickness of few  tens of Angstroms at the target surface, the number of reactive gas atoms was taken as the product of coverage, density and stoichiometry as proposed [8, 25]. The reactive gas atoms also follow a balance with respect to sputtering condition as proposed by the model [8]. The sputtering yield was calculated from ref [26].

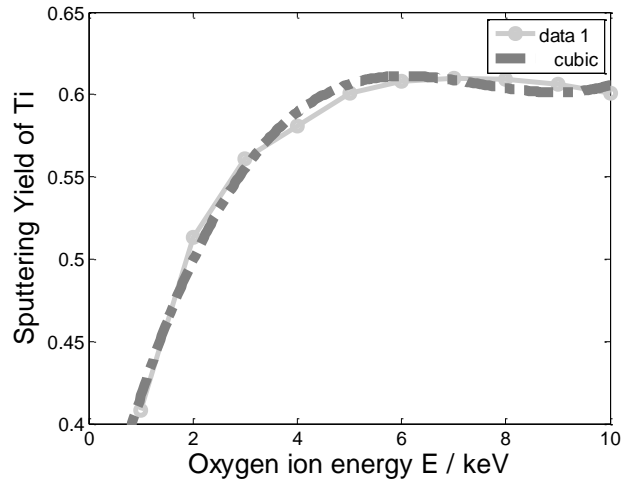

Fig 10: Variation of sputtering yield of Titanium with Oxygen ion energy and its cubic fit

The plot of sputtering yield of Ti with Oxygen ion energy along with a cubic fit is given in fig 10. Sticking coefficient, $\alpha$ is the rate of coverage with exposure. The coverage, $\theta$ is again the fraction of adsorbed atomic layer on the surface. Thus the value of $\theta = 1$ when one atomic layer of the adsorbate has formed on the surface. The surface can be either the target or the substrate but with different sticking coefficients. At the target surface the compound is formed by chemisorption of reactive gas. So (*1- (i))* is the fraction of the metal target where compound formation has not taken place. If f is the flux of reactive gas atoms and $\alpha_{tm}$ the sticking coefficient of the metal part of the target surface. Then $\alpha_{tm}$ (*1- (i))* f is the number of reactive gas atoms introduced to the surface [24].

The adsorption process on a substrate can be of three types. The first case is one where the *adlayer* forms on a substrate with constant $\alpha$ until a monolayer ($\theta = 1$) is formed. This type of growth is typical for metals on metals. The second type has characteristics of an increased $\alpha$ till one ML due to clustering of adatoms but a decrease further in $\alpha$ due to coalescence which is typical of metals getting deposited on alkali halides. For gas adsorbtion in metals like the case which is being discussed here, the adatoms occupy the unadsorbed sites and $\alpha$ found to fall exponentially as the coverage increases and attains saturation at $\theta=1$ [27]. So we can assign an equation $\theta = (1 - e^{-x})$ to the coverage and $S = e^{-x}$ to the sticking coefficient. The parameter "$x$" is stoichiometry of the compound being formed and is related to the slope of the plot.

After the first layer of compound has formed on the surface, sputtering of the compound layer starts which also exposes the target metal surface and subsequently further compound formation in the exposed region. The compound adatoms ejected per unit time can be expressed by $jx\theta$ where j is the density of ion impacts.

The rate of sputtering i.e. the amount of target materials sputtered per unit time is given by eqn 1, where M is molar weight of the target [kg/mol]; r is the density of the material [kg/m$^3$]; NA is the $6.02 \times 10^{26}$ 1/kmol; (Avogadro number); e is the $1.6 \times 10^{-19}$ As (electron charge); S is the sputtering yield (atom/ion) and jp: primary ion current density [A/m$^2$]. This rate divided by atomic diameter gives us the rate of sputtering in terms of atomic layers (AL) per second [15].

$$zt = M/(rNAe)\, S\, jp \qquad (1)$$

The ionic radius of the titanium (IV) ion is *0.745 Å* and that of the oxide ion is *1.26 Å*. The ratio of radii for the cation and anion is thus $r_+/r_- = 0.745/1.26 = 0.591$. Titanium has a radius of 1.47 Å. As Ti gets adsorbed by $O_2$ to form a TiOx compound layer, we assume the radius to be varying from 0.745 to 2 Å.The sputtering yield of metal from the compound layer was taken to be STic = 0.03 whereas the sputtering yield of the reactive gas from the compound layer was taken as $SO_2 =$ 0.09 as reported in ref [24]. Therefore we end up into two different sputtering rates: one for metal and other for the reactive gas. As the slowest step determines the rate of reaction, we can infer that the rate of TiOx deposition will be that of Ti in this circumstance.

The adatoms sputtered from the target are multiplied with exp (-d/L) to get rough estimate of the deposited atomic layers per second. L (cm) is the mean free path of the sputtered atom which is related to the sputtering pressure P (Torr) and the molecular diameter Da of the reactive gas (252 pm for $O_2$) $L = 2.303 \times 10^{-20} T / (PDa^2)$ where T is the deposition temperature and d (cm) is the distance traversed by the adatom for deposition which was taken as 30 cm considering the target the substrate distance [7, 28]. A fraction of 2.3540 atomic layers of TiOx gets deposited in 30s and $10^{-3}$ Torr pressure at a current density of 1 mA/cm².

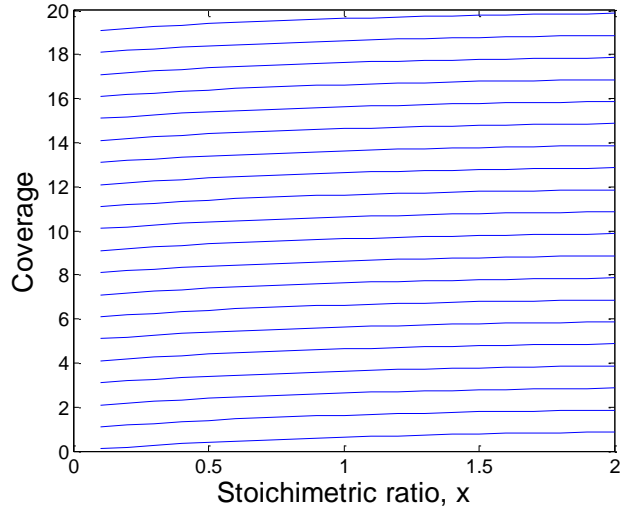

Fig 11: The variation of TiOx coverage with respected to stoichiometric ratio x at increasing atomic layer deposition.

About 21 atomic layers of oxygen sputtered from the compound layer at a sputtering yield of 0.09 are deposited on the substrate with the above mentioned conditions. These atomic layers will interact with the titanium metal atomic layers and form the TiOx coating. The rate of change of stoichiometric ratio is proportional to the change in the atomic layers /coating thickness with respect to adatoms exposure However as only 2.3540 atomic layers of TiOx is deposited, there is sufficient oxygen available to take the stoichiometric ratio to 2 making a $TiO_2$ film. The stoichiometric ratio x was varied from 0.1 till 2 corresponding to the formation of TiO $_{x=0.1}$ to $TiO_2$. The relation for coverage for monoatomic layers now gets changed to that for 2.354 atomic layers as given in eqn 2.

$$\theta = (N - e^{-x}) \qquad\qquad (2)$$

where N is the number of adlayers of TiOx. The variation of coverage of TiOx considering oxygen getting adsorbed on a previously deposited Ti layer is shown in fig 11. The plot closest to the x axis is for monolayer deposition which increases as we go upwards till 20 atomic layers. The deposition parameters were varied to get different atomic layers. The nature of variation for Cu or for any metals for that matter is shown in ref [7]. The nature was also same in this case. The decrease in atomic layer with pressure  is simply due to lower mean free path of gas atoms at higher pressure, the  computational as well as experimental evidence of which has also been reported earlier [7, 17, 29].

**Conclusions**

A model was developed for reactive sputtering of TiO$_x$ film in oxygen atmosphere. It was based on the nature of coverage with exposure of adatoms and the stoichiometry of the compound film. A variation of coverage with stoichiometry for different atomic layers was obtained.

This section is stored in repository [30]

## 4. Sputter Modeling of Titanium Nitride

The surface behaves different from bulk. It possesses different atomic structure and crystal structure. The electronic, transport as well as physical and chemical properties are quite different from bulk [27]. We need to study surface and try to improve its properties as it has tremendous technological importance. There are various methods to modify a surface: heat treatment, plasma surface modification and thin film deposition to name a few. Among all the surfaces, Ti surface is a significant one especially in areas where high strength components are requires like aerospace. In this communication focus has been given on TiN surfaces. TiN are materials with high chemical and thermal stability as well as good mechanical properties. They are used extensively in semiconductors as well as for protective and decorative purposes [31, 32]. A modeling has been shown for the growth mechanism of TiN films during reactive sputter deposition. This model has been developed for the epitaxial growth which is based on the atoms getting sputtered out from the target surface and ultimately being deposited on the substrate to form the film [7]. The rate of  sputtering i.e. the amount of target materials sputtered per unit time is given by eqn 1, where M is  molar weight of the target [kg/mol]; r is the density of the material [kg/m$^3$]; NA is the  $6.02 \times 10^{26}$ 1/kmol; (Avogadro number); e is the $1.6 \times 10^{-19}$ As (electron charge); S is the sputtering yield (atom/ion) and  jp: primary ion current density [A/m²]. This rate divided by atomic diameter gives us the rate of sputtering in terms of atomic layers (AL) per second [15]

$$zt = M/(rNAe) \, S \, jp \qquad (1)$$

There is a preferential sputtering of Nitrogen if a TiN target is used. Also the partial sputtering yields as well as the sticking coefficients of Ti and N are different. The sputtering yield of Ti was taken from 0.1 to 1.0 as per ref [31]. The radius of Ti is 147 pm and that of N is 56 pm. Therefore the radius of TiN was taken as 200 pm as a rough estimate.

The adatoms sputtered from the target are multiplied with exp (-d/L) to get rough estimate of the deposited atomic layers per second. L (cm) is the mean free path of the sputtered atom which is related to the sputtering pressure P (Torr) and the molecular diameter Da of the reactive gas (252 pm for $O_2$) $L = 2.303 \times 10^{-20} T / (PDa^2)$ where T is the deposition temperature and d (cm) is the distance traversed by the adatom for deposition which was taken as 30 cm considering the target the substrate distance [7, 15].

A ratio was given in ref [31] as the ratio of the partial yields of Ti and N as given below in eqn 2. The ratio q depends on the composition of the target and energy of the incident ion.

$$q = \frac{s_N}{s_{Ti}} \qquad (2)$$

The atomic flux of N/Ti plays a key role in the growth of TiN films [33]. The change in composition during the sputtering process was not considered which has been tried to inculcate here. So instead of q we introduce here the rate of change of q w.r.t coverage (= dq/dθ). The rate decreases from the initial value due to preferential sputtering of N and the target is deficient of N. Based on our previous publication, we can assign an equation $\theta = (1 - e^{-x})$ to the coverage and $S = e^{-x}$ to the sticking coefficient. The parameter "$x$" is stoichiometry of the compound being formed [30]. The variation of partial yield and derivative of partial yield with coverage is shown in fig 12

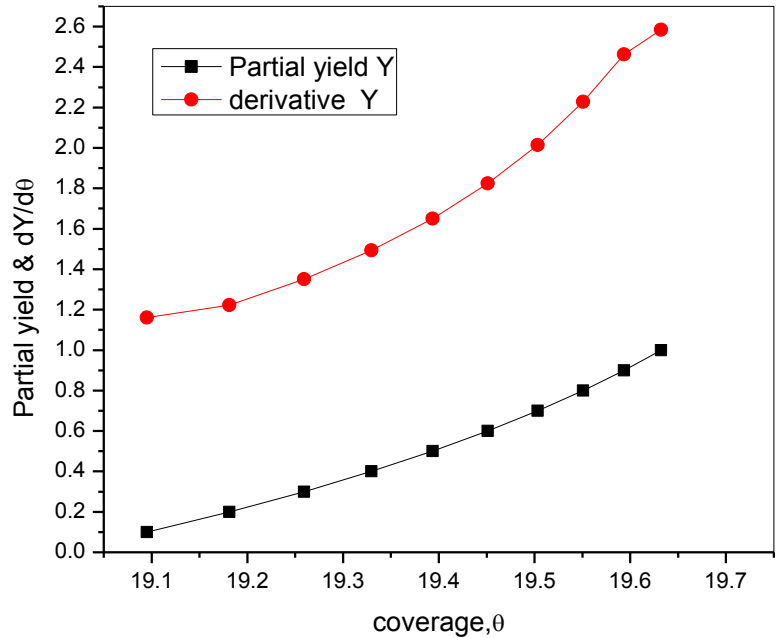

Fig 12: Variation of partial yield and derivative of partial yield with coverage

The variation of deposition pressure caused difference in deposition thickness of the TiN films as shown in fig 13. Increase in vacuum caused a decrease in deposition which is quite different phenomenon. From fig 14 a deposition of upto 200 nm was observed on changing the deposition time to 100s.

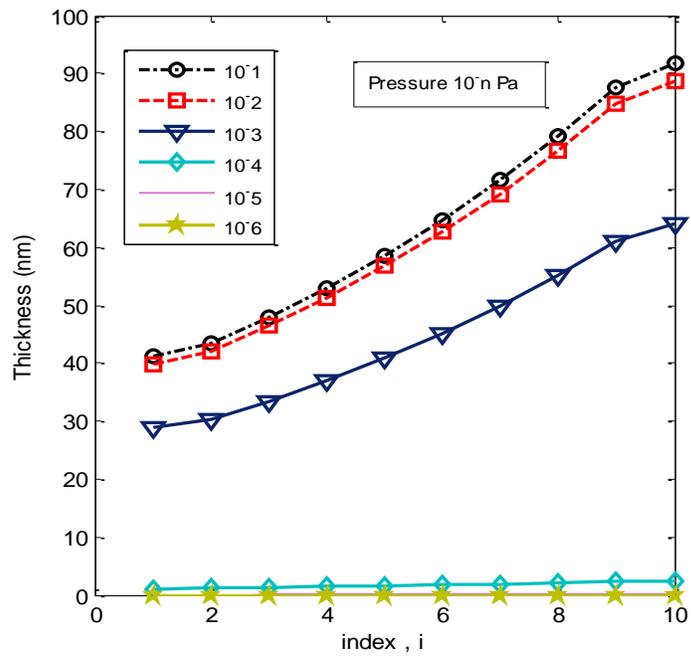

Fig 13: Simulated TiN of various thicknesses deposited at different pressures

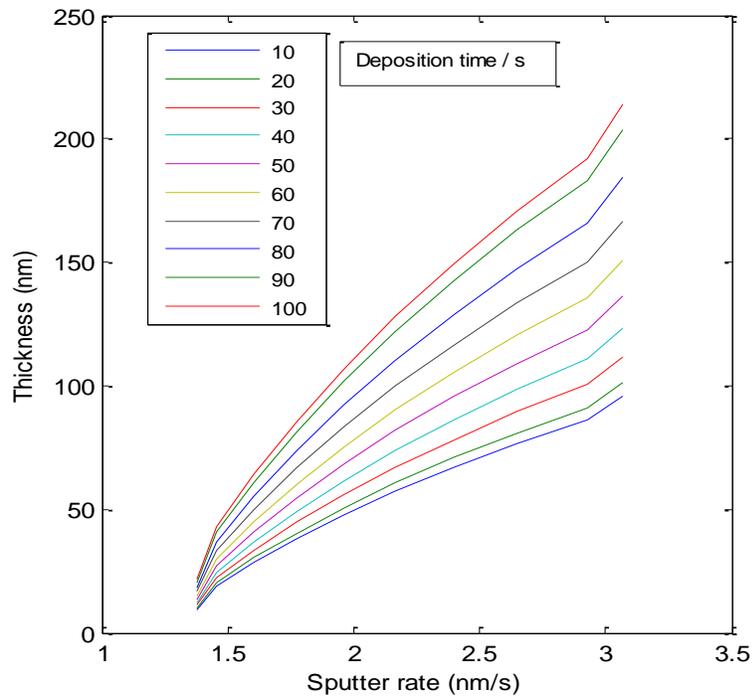

Fig 14: TiN film thickness variation with sputter rate at different deposition time.

## Conclusions

A model was developed for sputtering of TiN film. An increase in the rate of partial yield with coverage was observed. The variation of deposition pressure and deposition time was obtained. This section is again stored in a repository [34]

## Reference

1. W. J. de Haase and P. M. van Alphen, Proc. Acad. Sci.33, 1106 (1930).
2. Y. F. Ogrin, V. N. Lutskii, and M. I. Elinson, JETP Lett.7, 71 (1966).
3. A. A. Abrikosov., Europhys. Lett.49, 789 (2000).
4. W. Buckel and R. Hilsch, Z. Phys.138, 109 (1954).
5. K. Behnia, L. Balicas, and Y. Kopelevich, Science 317, 1729 (2007)

6. Naresh T. Deoli, Lucas C. Phinney, Duncan L. Weathers, Nuclear Instruments and Methods in Physics Research B 332 (2014) 286–289

7. AS Bhattacharyya, S Kumar, S Jana, PY Kommu, K Gaurav, S Prabha, VS.Kujur, P.Bharadwaj, Sputter based Epitaxial Growth and Modeling of Cu/Si Thin Films, *Int. J. Thin. Fil. Sci. Tec* 4 (3), 2015,173-177 and in arXiv:1508.04920 [cond-mat.mtrl-sci] 2015

8. A.S.Bhattacharyya, Sweta Prabha, Vidya S. Kujur, P. Bharadwaj, Computation modeling of epitaxial deposition of Tungsten thin films J. Nano science Engineering & Trends. *2(1), 2016*,1-5

9. Shunhao Xiao, Dahai Wei, and Xiaofeng Jin, Bi(111) Thin Film with Insulating Interior but Metallic Surfaces Phys. Rev. Lett. 109, 166805 – Published 17 October 2012

10. Kai Zhu, Lin Wu, Xinxin Gong, Shunhao Xiao, Shiyan Li, Xiaofeng Jin, Mengyu Yao, Dong Qian, Meng Wu, Ji Feng, Qian Niu, Fernando de Juan, Dung-Hai Lee arXiv:1403.0066 [cond-mat.mes-hall] 2014

11. Wang D, Chen L, Liu H, Wang X, Cui G, Zhang P, Zhao D, Ji S. Topological states modulation of Bi and Sb thin films by atomic adsorption Phys Chem Chem Phys. 2015 Feb 7;17(5):3577-83. doi: 10.1039/c4cp04502e.

12. Justin Kelly Magnetotransport of Topological Insulators: Bismuth Selenide and Bismuth Telluride 2011 NSF/REU Program, Physics Department, University of Notre Dame

13. Ziga Kos, Topological Insulators, University of Ljubljana, June 24, 2013

14. A.S.Bhattacharyya, R. Praveen Kumar, Rishideo Kumar, Vikrant Raj, *Bismuth Thin Films: Polar Angle and Ion Fluence*, 2016, eprint: Condensed Matter > viXra:1603.0420

15. Useful information and facts about sputtering, http://www.specs.de/cms/upload/PDFs/IQE11-35/sputter-info.pdf

16. Sweta Prabha, Vidya S. Kujur, P.Bharadwaj, B.Tech project report, Nanotechnology, Central University of Jharkhand, 2015

17. A.S. Bhattacharyya , Bragg Curve in Sputtering, DOI: 10.13140 / RG. 2.1.2735. 5366.

18. W.D. Sproul, D.J. Christie, D.C. Carter, Control of reactive sputtering processes, Thin Solid Films 491 (2005) 1 – 17.

19. A. Okamoto, T. Serikawa, Thin Solid Films 137 (1986) 143.

20. S. Schiller, U. Heisig, K. Steinfelder, J. Strumpfel, Thin Solid Films 96 (1982) 235.

21. William D. Sproul, James R. Tomashek, U.S. Patent 4,428,811, January 31, 1984.

22. J. Affinito, R.R. Parsons, J. Vac. Sci. Technol., A 2 (1984) 1275

23. Tomas Nyberg, Soren Berg, PCT Patent Application WO 03/006703 A1, 23 January 2003.

24. T. Kubart, O. Kappertz, T. Nyberg, S. Berg, Dynamic behaviour of the reactive sputtering process Thin Solid Films 515 (2006) 421 – 424

25. D. Rosen, I. Katardjlev, S. Berg, W. Moller, Nucl. Instrum. Methods Phys. Res., B Beam Interact. Mater. Atoms B 228 (2005) 193.

26. http://www.iap.tuwien.ac.at/www/surface/sputteryield

27. M.Prutton, Introduction to Surface Physics, Clarendon Press. Oxford, 160

28. Vipin Chawla, R. Jayaganthan, A.K. Chawla, Ramesh Chandra, Journal of materials processing technology; **209** (2009) 3444–3451

29. A.S.Bhattacharyya, S.K.Mishra, S.Mukherjee J. of Vac. Sc. Technol. A 28 (2010) 505-509.

30. A. S. Bhattacharyya, K Gaurav, P Kommu, Compound TiOx film growth dynamics in Sputtering, DOI: 10.13140/RG.2.1.5077.808

31. Rajiv Ranjan, J. P. Allain, M. R. Hendricks, and D. N. Ruzic J. Vac. Sci. Technol. A 19.3., 2001,1004-1007

32. Nikhil K. Ponon, Daniel J.R. Appleby, Erhan Arac, P.J. King, Srinivas Ganti,

Kelvin S.K. Kwa, Anthony O'Neill, Effect of deposition conditions and post deposition anneal on reactively sputtered titanium nitride thin films Thin Solid Films 578 (2015) 31–37

33. S Mahieu· D Depla , Reactive sputter deposition of TiN layers: modelling the growth by characterization of particle fluxes towards the substrate, Journal of Physics D: Applied Physics,42, 053002

34. A.S.Bhattacharyya, R. Praveen Kumar, Titanium Nitride: Sputter Modeling, e-print viXra: 1510.0331. 2015